\documentclass{article}
\usepackage{preamble}

\title{{\huge
Socioeconomic factors of national representation in the global film festival circuit: skewed toward the large and wealthy, but small countries can beat the odds
}}
\author{Andres Karjus\textsuperscript{1,2}, Vejune Zemaityte\textsuperscript{1} \\
\textsuperscript{1}Tallinn University, \textsuperscript{2}Estonian Business School }
\date{\vspace{-0.8cm}}

\begin{document}

\maketitle
\begin{abstract}
    This study analyzes how economic, demographic, and geographic factors predict the representation of different countries in the global film festival circuit. It relies on the combination of several open-access databases, including festival programming information from the Cinando platform of the Cannes Film Market. The dataset consists of over {20,000} unique films from almost 600 festivals across the world over a decade, a total of more than {30,000} 
    film-festival entries.
    It is shown that while films from large affluent countries indeed dominate the festival screen, the bias is nevertheless not fully proportional to the large demographic and economic worldwide disparities and that several smaller countries perform better than expected. Further computational simulations demonstrate how much including films from smaller countries contributes to cultural diversity, and how countries vary in cultural "trade balance" dynamics, revealing differences between net exporters and importers of festival films. This research underscores the importance of representation in film festivals and the public value of increasing cultural diversity. The data-driven insights and quantitative approaches to festival programming and cultural event analytics are hoped to be useful for both the academic community as well as film festival organizers and policymakers aiming to foster more inclusive and diverse cultural landscapes.
\end{abstract}

\section{Introduction}

Film festivals are a vital component of the film industry, occurring in various forms worldwide, ranging from major industry hubs like Cannes, Sundance, or the Berlinale, to regional and smaller national events.
These festivals differ in their budget, audience size, specialization, acceptance rates, and diversity, but form a well-connected global network in terms of the flow of recurring films, filmmakers, and industry stakeholders \parencite{krainhofer_mapping_2018,loist_film_2016, zemaityte_quantifying_2024}. 
They offer filmmakers a venue to showcase and market their productions, which can determine their further success \parencite[][]{stringer_global_2001}. 
Festivals also promote cultural interactions and dialogue between makers and audiences, public and industry interests, and generate public value \parencite[in the sense of][]{benington_public_2011,oregan_past_2022,mazzucato_creating_2020} 
to societies and public spheres, by showcasing culturally varied programs that broaden public access to films and foster international cultural learning and exchange \parencite{harbord_film_2002,ruling_film_2010,de_valck_film_2007,roy_telling_2014,de_valck_fostering_2016,mair_role_2020,diestro-dopido_film_2021}. 
Increasing diversity in cultural landscapes, including by platforming otherwise low capacity or underrepresented voices and potentially novel perspectives, can in turn foster societal resilience and innovation \parencite{farrell_understanding_2016,burgess_capturing_2020,hartley_digital_2020}. Cultural exports including films can also act as soft power \parencite{guan_winning_2023}.

While balancing between various pressures --- like marketability, artistic aspirations, activism, diversity, and audience choice --- can be challenging in terms of curation \parencite{bosma_film_2015}, power over programming means festivals can also bypass traditional distribution limitations and address topical issues \parencite{de_valck_introduction_2016,elsaesser_european_2005}. 
The role of film festivals also extends beyond exhibition and cultural learning. Like other cultural festivals \parencite[][]{wilson_expanding_2017}, they boost local economies, create financial and symbolic value for host countries and regions via public funding and subsidies, create employment opportunities, and increase revenues via tourism
\parencite{monson_economic_2023,locarno_film_festival_annual_2022,grunwell_film_2008,harbord_film_2002,kendall_film_2021,kostopoulou_cultural_2013}.

\begin{figure}[htb]
	\noindent
	\includegraphics[width=\columnwidth]{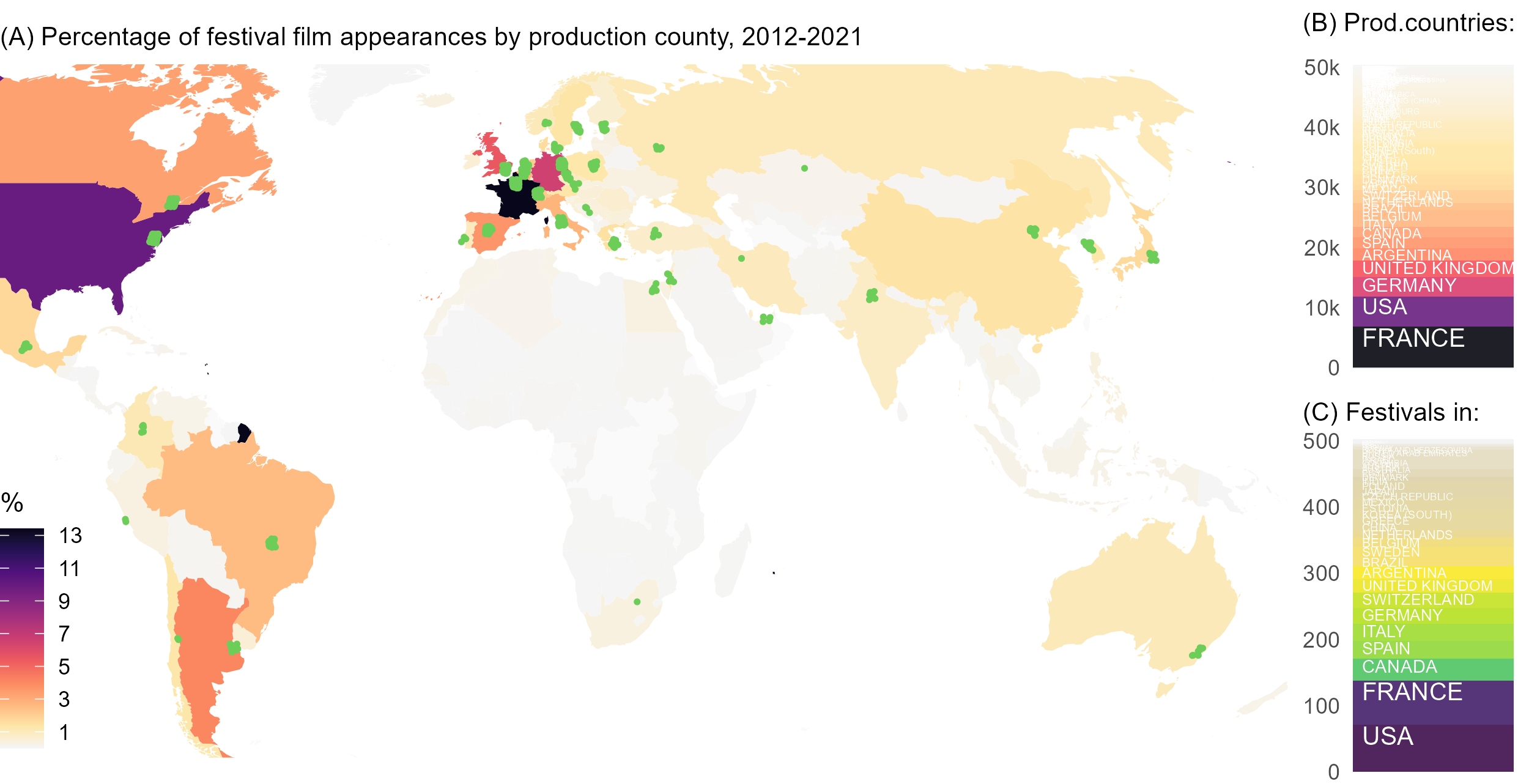}
	\caption{Mapping the film festivals and production countries across the globe. Panel A shows the volume of 
 film-festival pair entries in the database (indicating one or more screenings at a given festival)
 in the circuit from a given country, and festival locations as green dots (placed over country capitals). Panels B and C are stacked bar charts, showing the cumulative number of pairs and hosted events by country, respectively. Both stacks are dominated by the USA and France, followed by a long tail of less prominent production and host countries. The smaller countries with lesser shares are explored on subsequent graphs.
	}\label{fig_map}
\end{figure}

But the festival circuit is not a level playing field. Countries that are smaller, less affluent or whose language is not among the few international lingua francas --- naturally face various challenges when it comes to the production as well as (local and global) distribution of cultural products, including films \parencite{hjort_cinema_2007,martin-jones_uruguay_2013,macpherson_is_2010,fu_examining_2010}.
The dynamics of programming content from large global producers and small countries in festivals, cinema, and television, including questions around the value of cultural diversity, are therefore an active research area \parencite[see e.g.]{ibrus_what_2014,sand_small_2019,felix-jager_effects_2020,navarro_local_2022,zemaityte_quantifying_2024,ibrus_quantifying_2023,zemaityte_peripheral_nodate}.
This contribution makes use of several sources of open data, including a refined version of a snapshot of the Cinando database of international film festivals \parencite[][]{zemaityte_cinando_2023,zemaityte_quantifying_2024}, further described in Materials and methods below. A first glance at the data indicates that the global film festival circuit is rather biased towards large, wealthy nations (Figure \ref{fig_map}). Almost a quarter of the films in the circuit in 2012-2021 listed either France or the USA as the production or co-production country, and over a quarter of the festivals took place in these two countries (see Fig. \ref{fig_map}.C, and the green dots on Fig. \ref{fig_map}.A). The database includes films from 163 countries, but half the films in that period were made in or involved one of the top 8 most productive countries. 

However, smaller countries likely have fewer filmmakers and less prosperous nations have fewer resources to invest in film production. Iceland cannot be expected to match the volume of American film production; Switzerland and Latvia are physically closer to more international festival locations than Malaysia or Uganda. Films from the UK or Australia 
may be easier to follow culturally and language-wise
in many places due to being in English. Filmmakers or stakeholders in countries distant from the current hot spots may find fewer opportunities for travel to showcase and advertise their products. India is a large film producer, but not all Bollywood films might make it to the mostly West-centric festivals nor be aimed at them. First-world countries simply have more resources to drive their film industry and local cinema-going audiences with higher disposable incomes. Yet if economic and demographic disparities lead to unequal representation, it can undermine the potential for film festivals to democratize access to diverse cultural expressions and public value available to global audiences.

This short contribution attempts to shed light on these dynamics and fill a gap in film festival research literature by directly measuring national representation at festivals, and comparing this to countries' economic and demographic factors, as well as providing a quantitative cultural analytics of the flows of festival films between countries of the world.
Until recently, film festival research has remained mostly theoretical \parencite[e.g.][]{de_valck_film_2007,loist_film_2016,wilson_expanding_2017}, qualitative or participant experience focused \parencite{peirano_mapping_2020,mair_role_2020,sand_small_2019}, or based on smaller data samples \parencite{ehrich_film_2022}.
Following more recent research \parencite{zemaityte_quantifying_2024,zemaityte_peripheral_nodate}, this contribution study aims to leverage big data, systematic quantitative analysis, and predictive modeling.
Particular attention is paid here to small countries, how they fit into the festival ecosystem, and how much diversity and therefore public value they add to the circuit. 
In summary, the following research questions are addressed:
\begin{enumerate}
    \item To what extent is the festival circuit skewed or biased towards populous and wealthy countries?
    \item To what extent are factors like relative wealth and population size predictive of national representation in the film festival circuit? What characterizes countries that perform better or worse than expected? 
    \item  How and how much do countries differ in terms of the cultural "trade balance" of festival films?
    \item Is the cultural diversity of the festival landscape affected by the inclusion or absence of smaller and less prosperous countries?
\end{enumerate}

\section{Methods and materials}

\subsection{Film festival data}

This research makes use of three large open-access datasets and two additional sources of information. The primary source is the Cinando dataset \parencite{zemaityte_cinando_2023}, published along with \textcite{zemaityte_quantifying_2024} and further explored in \textcite{zemaityte_peripheral_nodate}. It is a cleaned version of a static snapshot of the proprietary relational database underlying the namesake platform operated by the Cannes Film Market (also known as Marché du Film - Festival de Cannes). The 2012-2021 subset used for this study is operationalized as a set of 50,408 data points, where each point is the 
listing of a film at a film festival (the database does not specify how many times a listed film was screened at a given festival, but at least once can be assumed). 
If there are multiple co-production countries, then the listing is repeated for each to include the country metadata (but this is taken into account in the analyses). Ignoring co-productions, there are 34,535 
such film-festival data points, 
in turn reflecting high levels of co-production \parencite[cf.][]{bondebjerg_transnational_2016,parc_understanding_2020}. 
There are 22,167 unique films in this sample (approximately; see below), 592 festivals in 129 event series, 148 
primary production countries (listed first in the ordered producer variable in the database)
and 163 countries in total, i.e. 84\% of the world's countries occur at least once.

While the Cinando snapshot used here is currently one of if not the most extensive global dataset available on the programming of international film festivals, it comes with limitations stemming from its initial European origins, having started as a service solely for the Cannes Film Festival. Using it has arguably later become an industry standard, however, reflected by the wide range of festival events recorded there and correlating with UNESCO production estimates (see below). The potential biases include possible under-representation of non-European events and industries, and analyses based on it risk suffering from "survivorship bias", as the database or at least the version available here does not record rejections by the programming committees, only programs of successfully selected films. The initial years of the database are also sparse, reflecting its gradual uptake by the industry. \textcite{zemaityte_quantifying_2024} used a subset of data between 2009 and 2021 for most, and from 2012 for more conservative analyses. The latter approach is taken here for aggregated data, as including 2009-2011 risks biasing the results towards France and French films by showing larger shares and counts than would be representative of the subsequent decade.

Most of the data appears to be entered by platform users, requiring extensive data cleaning described in \textcite{zemaityte_quantifying_2024}. One shortcoming that remains is the identification of individual films, as films have been often entered into the database multiple times using different ID values. Here, films are further identified using the title and the reported production year, but both of those values are also known to occasionally change during film production, and some films have short single-word titles that can easily be confused. Another shortcoming is festival dates, limited to years in the dataset, which in turn constrains temporal analyses. The cultural flows section in the Results only considers direct links between film production and festival host countries but not the journeys of films through the festival circuit, including getting selected for a prestigious festival that may affect subsequent selections. This would require data with more precise festival dates than what is currently available.

Another technical detail concerns co-productions. The database does include an order variable for metadata, including production countries, but this is difficult to interpret systematically and is ignored here as in \textcite{zemaityte_quantifying_2024}. The existence of multiple producers is accounted for the the analyses however, but doing so requires choices about aggregation to match the goals of a given analysis. In the study on representation below, each listed country is counted as one full entry, whereas in the analyses on cultural flows and diversity, they are weighted, so that for multiple co-producers $n$, the weights equal $1/n$ i.e. sum to one (see Results section).

The available data only includes programs, i.e. already selected films at festivals --- there is no information on acceptance rates and rejections, which would be interesting to investigate of course if available. What can still be modeled is aggregated counts of 
film-festival pairs
per country, which is done below. The current count approach ignores individual festivals, and it is not possible to model effects such as festival types and distance of the festival from the production country, nor to control for the festival year (the period is just under 10 years however). There are of course other ways to aggregate, but each has its downsides. The percentage of countries in a festival is one option, but that would give undue weight to small festivals. Binomial representation, 
i.e. just counting whether a country is present in a festival or year or not,
would enable modeling more detailed metadata, but that would make the results equivalent between having a single film at a festival from a given country and films from that country filling the entire festival program (and be largely a dataset of zeroes, especially if done per year, given that many countries in the dataset only occur a few times). 

In the linear regression modeling below, counts on a logarithmic scale are used as the response variable. Another option would be a Poisson regression or a negative binomial model, given the large discrepancy between a few large counts and a long tail of small values. The log-linear approach lends itself to a more convenient interpretation, especially given that all other variables are similarly power-law distributed and also on the log scale. 
For the representation and bias analysis, the FIAPF or International Federation of Film Producers Associations accreditation list is used as additional categorization \parencite{fiapf_accredited_2023}. This list represents the so-called "A-list" top competitive feature film festivals at the center of the festival industry, widely recognized by filmmakers, although it does not include all large and otherwise significant event series such as Sundance. Most of the festival series in the dataset, 115 out of 129, are B-listers however.

\subsection{Socioeconomic and film production data}

The two other datasets are the World Bank (WB) database \parencite{world_bank_group_world_2024}, which is used to source information on population and gross domestic product (GDP) as in \textcite{zemaityte_peripheral_nodate}, and the UNESCO Institute of Statistics database on the counts of feature films produced by country \parencite[][]{uis_unesco_2024}. 
Both were aligned with the Cinando dataset by country name (manually correcting names where there was a mismatch in naming formats). Where data for a year was missing for any country, linear interpolation was used to predict the values from adjacent present years.

The UIS production numbers are not directly comparable to the Cinando festival programming data, even if the latter were transformed into unique film counts. The data collection methods are clearly different, and UIS counts only feature films, whereas festivals feature various formats including short films. Therefore, instead of measuring over or under-representation in absolute numbers, a correlation between the magnitudes is instead provided in the Results section. This relationship could be further analyzed in future research.

These data from the additional databases were added by production country to the final dataset, but aligned by the year of the festival event, not the year of film production. This may be a debatable choice but hopefully leads to more informative results. One argument is the existence of retrospective festivals or program segments. As an extreme example, the socioeconomic situation of the United Kingdom in 1927 probably has little bearing on the screening of the film "The Ring" from that year in Cannes in 2012. The economic situation of a country could itself be a factor, however, when it comes to representation in a given year in the festival circuit. Furthermore, films typically stay in the circuit only a few years after production anyway, including due to rules of admission in many festivals  \parencite{zemaityte_quantifying_2024}. In the Cinando subset, the average difference between the year of production and festival occurrence is just 0.7 years (both variables are at yearly resolution, so this is a somewhat rough estimate).

These should be considered however only rough estimates of more interesting target measures, which would be e.g. the number of filmmakers in a country
and its media market size, 
and the financial means, revenue, and subsidies of a country's film industry. Total population is a proxy, but of course, it cannot be expected that the number of filmmakers is proportional in every country.  
GDP is the total value of produced goods and services produced but says nothing about wealth distribution or investment in creative industries. Since GDP and population typically correlate, the measure used in the analyses here is GDP per capita, i.e. divided by the population estimate of a given year, as a rough proxy of relative prosperity.
These variables are averaged for the analyses below for the 9-year observation period, weighted by the number of 
film-festival pairs
in a year: this way the aggregate for each country is most representative of when it was most present in the circuit.

While the preceding literature discussed in the Introduction has discussed and compared small and large countries, using continuous population and GDP variables here means there is no need to categorize or bin countries into any arbitrary discrete categories such as small-medium-large, which in turn could easily bias analyses down the line. Unlike \textcite{zemaityte_peripheral_nodate}, which this analysis otherwise complements, the models here of representation, bias, and diversity all use the continuous scales of the aforementioned variables.
A country as such is not necessarily an ideal unit either, as many countries may include regions with different population levels, affluence, or access to arts education. Here, country and nation are used interchangeably for simplicity, but many countries of course consist of multiple ethnic and cultural groups, which again may be associated with differential access to resources.
Another simplification concerns location and distance. The Cinando dataset only includes production and festival host information at the country level, so capital city locations and inter-distances are used where needed (including the map in Fig. \ref{fig_map}.A and the regression model in the results).

\subsection{Thematic and linguistic diversity}

The diversity section in the Results below follows the method of diversity calculation and the open source codebase of \textcite{zemaityte_quantifying_2024}, who argue against using discrete labels for calculating diversity when the distances between categories are not equidistant \parencite[cf.][]{moreau_cultural_2004}. This is relevant in data like thematic metadata tags such as drama or horror, where some genres are inherently more similar than others. 
Languages of the world also vary in similarity, and the delineation of language, variety, or dialect is often more political than anything else. For example, a festival featuring films in Dutch, Flemish, and German would be as diverse on paper as one featuring Japanese, Zulu, and Finnish, if one were to simply count the labels. Their proposed solution used latent spaces, induced directly from the tag co-occurrence data for genres (analogously to word embeddings), and an externally sourced language typology vectors dataset for linguistic similarity \parencite{malaviya_learning_2017}. 
The set of 41 possible thematic or genre tags in the Cinando dataset is admittedly not ideal, consisting of broad genres (drama, comedy, documentary, etc.), but also markers like animation or book adaption, and production-related tags like first film, female producer and Bollywood. Here as in the aforementioned paper, the thematic vectors are averaged for each film, i.e. the latent "meaning" of a film is allowed to be a composite of its assigned metadata tag semantics.

The continuous spaces then enable the calculation of diversity as the average distance from a global mean vector (analogously to mean absolute deviation or MAD). 
The mean is normalized by the maximum possible distance in a given space, yielding a metric in the $[0,1]$ range.
The intuition is that a distribution tight around the (multidimensional) average indicates low diversity --- everything is close to the average. A high mean distance from the average indicates a wider spread and therefore higher diversity. 
Diversity can be conceptualized in various ways: this operationalization is meant to capture the macro-level, ecosystem-wide, or "external" diversity \parencite[following][]{mcquail_diversity_1983,zemaityte_quantifying_2024}.

As with any metric based on means, including MAD, this naturally assumes a (here multivariate) normal distribution. Given the normalization, a value of 1 is practically only possible in a bimodal distribution, e.g. a system consisting of only two of the most different but frequent categories --- which is probably no longer an intuitive representation of diversity, nor a likely policy goal. This limitation of the metric as it approaches its maximum should be kept in mind, but is not found to be an issue in the analyses below.

Finally, the simple diversity simulation described in the Results that uses the diversity measures is based on the concept of bootstrapping. For each subset of e.g. population level, the Cinando dataset of N entries is filtered according to the criterion, and then a new set of N entries is created using sampling with replacement from the filtered set. Therefore the samples, albeit filtered, are roughly proportional to the current representation shares of the countries, and the averages and confidence intervals are always calculated on the exact same sample size. To ensure reliability, this sampling procedure is repeated 100 times for each subset and averaged. When filtering, only films where all co-production countries match a criterion in the given year are kept (e.g. an Iceland-USA co-production would be left out given a sample with a population threshold of "$>300\textrm{M}$"). 
In other words, this approach measures what the festival circuit would look like if it comprised only countries matching a certain criterion, e.g. only above or below a certain size, and allows estimating how much diversity smaller or bigger countries contribute, based on the results without them. Another approach would be to do iterated weighted sampling with weights (directly and inversely) proportional to the population and GDP variables; this could be explored in future research.

\section{Results}

This section is organized by the research questions proposed in the Introduction, starting with the question of bias, followed by statistical modeling of representation affecting factors in the film festival circuit, modeling of cultural flows, and finally an assessment of estimated diversity contributions by countries of different socioeconomic profiles.

\subsection{Representational balance and bias in the film festival circuit}

\begin{figure}[tbh  ]
	\noindent
	\includegraphics[width=\columnwidth]{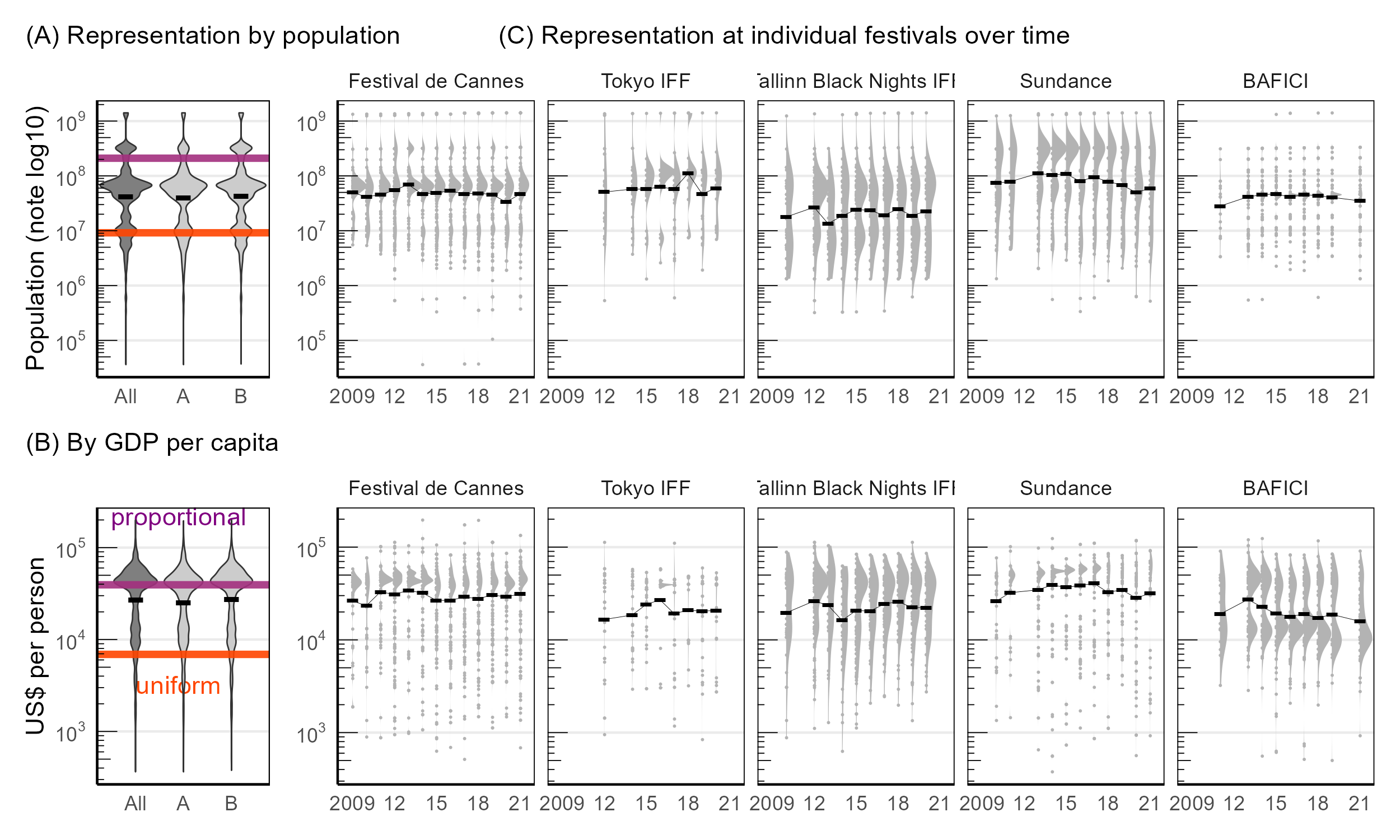}
	\caption{Representational equality in the global film festival circuit. Panels A and B on the left depict the circuit distributions (densities) and averages (black bars) in 2012-2021 for population and GDP per capita of production countries (each underlying data point is a 
 festival listing of a film from
from a given country), and the two models of balance as horizontal lines (see text for discussion). Note that the axes are logarithmic, and the bars are means of log-scaled values. The small panels in C, aligned with the axes of panels A and B, illustrate these values over time for a selection of individual festivals; the first three are FIAPF-accredited competitive event series, while Sundance and BACIFI (Buenos Aires International Independent Film Festival) are not. The graph shows most festivals stay relatively stable in terms of the average population size and affluence of production countries of films that get selected.
	}\label{fig_violins}
\end{figure}

\begin{figure}[htb]
	\noindent
	\includegraphics[width=\columnwidth]{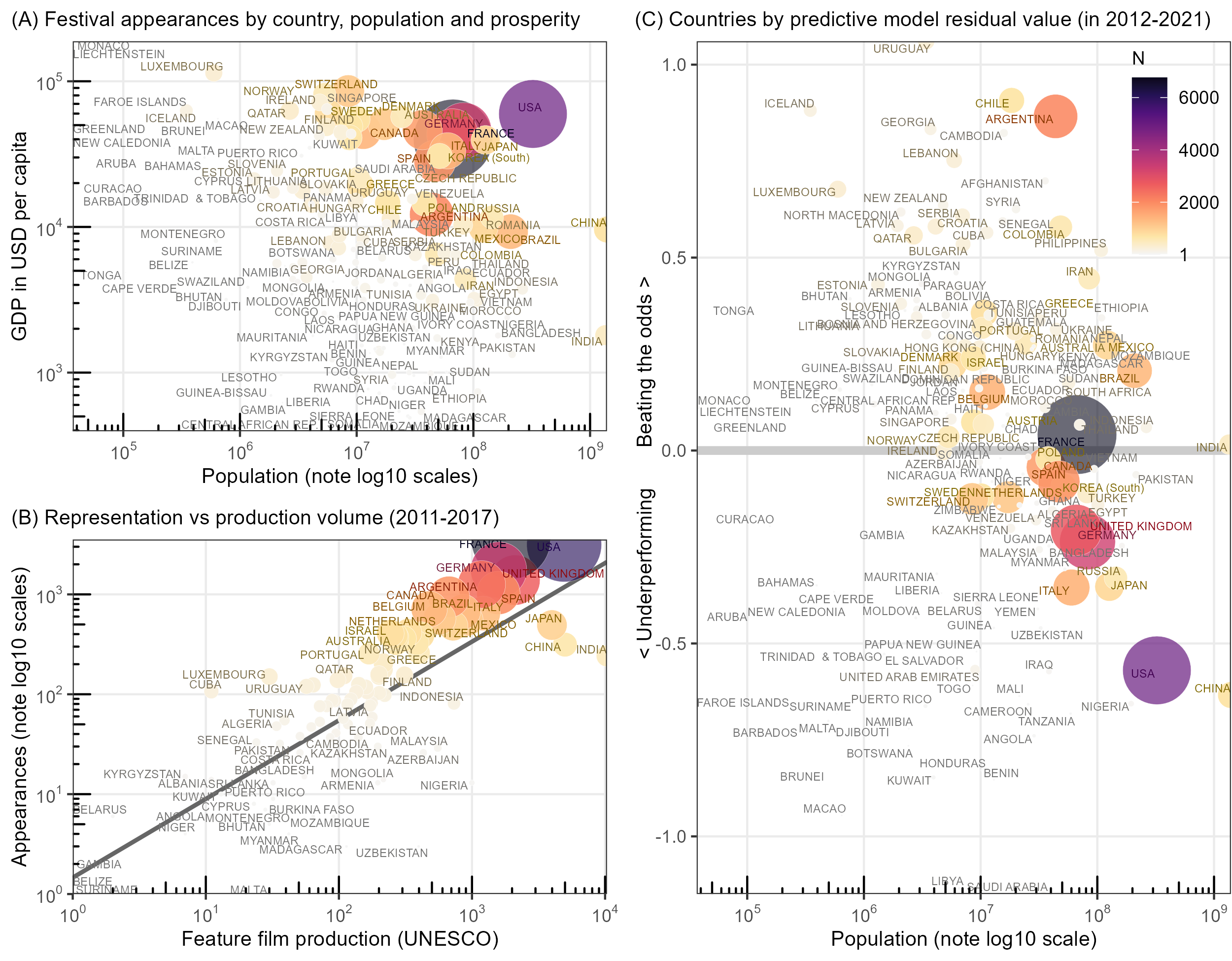}
	\caption{Three perspectives of festival film producing countries. Panel A has all the countries arranged by population and GDP per capita, with size indicating the number of
 film-festival pairs.
 As expected, the optimum is a wealthy economy and large population, while large African and Asian countries remain very much underrepresented. Panel B compares the number of 
 film-festival pairs
in the Cinando dataset and the UIS feature film production estimates in the overlapping period, showing they roughly correlate. Panel C shows the population on the x-axis and the residuals from the representation prediction model on the vertical: countries above the zero line occur more often than would be expected on average, while those below do worse given their population, GDP per capita, and other factors discussed in the text. Note that the Cinando dataset counts are based on 2012-2021 data on panels A and C, and 2011-2017 on panel B, to match the UIS data used in the latter.
	}\label{fig_gdppop}
\end{figure}

Quantifying balance bias depends on the unit of analysis, and a potentially ideological question of what counts as balance, equality, or equity. The result presented here is rather agnostic, offering two perspectives on balance across two variables. As discussed in the Methods section, the unit is the (co)production country of a film programmed at a festival; multiple appearances of a film at different festivals are counted as separate data points. Figure \ref{fig_violins}.A and B show the distribution of country appearances by their respective Population and GDP per capita (according to the WB). Both are on logarithmic scales, given widely known Pareto distributions of both variables. The same data is shown by individual countries in Figure \ref{fig_gdppop}.A.
The means (black bars) are therefore geometric rather than arithmetic averages.
The figure is further split by festival accreditation categories: all festivals, the prestigious competitive A-listers, and the rest (B-category) festivals. There do not appear to be particularly large differences between them on average, however. 
There is also not much difference in regions of the world, as most regions contain both small and large countries (see the complementary graph showing distributions by region in the Supplementary Materials appendix).
Figure \ref{fig_violins}.C shows the decadal trajectories of three A-list and two B-list festivals, demonstrating relative stability in the two variables over time. The Tallinn Black Nights International Film Festival is an A-lister, yet arguably smaller and less prestigious than e.g. Cannes or Sundance --- the latter itself a B-list yet widely known and influential festival series. Black Nights, taking place in the 1.3-million country of Estonia, also appears to program films on average from smaller and less prosperous countries compared to the others.

The two models of balance are labeled "uniform" and "proportional" \parencite{kedar_are_2016,kurz_democratic_2017}. The uniform expectation is calculated as the mean value of a simulated distribution of appearances where each country in the world (that has participated in the festival circuit according to Cinando), regardless of size and prosperity, gets to show the same number of films. These values are below the real-world means, again related to the power law: there are simply fewer very large or prosperous countries and more smaller ones. This model, while not necessarily the most realistic, demonstrates what (the average of) a festival circuit, blindly balanced by countries as units, would look like (or visually: where the black mean bar would be if the representation was uniform in this manner).

The second, proportional model is operationalized as the mean of a distribution where the number of festival appearances per country is perfectly proportional to their respective Population (Figure \ref{fig_violins}.A) and GDP per capita (B). The purple line shows where the mean (black bar) would be in this scenario. This illustrates a possible world where each country trains filmmakers and
hosts film production and exhibition industries
commensurate with their population size,
or produces films proportionally to their relative GDP. Here these factors are modeled separately for simplicity; their interaction is explored below. These values are higher than the real-world averages for both variables, demonstrating that the circuit is, from this perspective, indeed not as biased towards the larger and wealthier as it could potentially be. However, if one were to imagine a perfectly population-proportional international festival of say 100 films, then about 37 should be from China or India, 4 from the USA, and the rest of the 59 slots split between the rest of the 163 countries in the circuit. Depending on one's balancing goal, proportionality, diversity, and other goals should likely be taken into account to some extent.

The answer to RQ1 is therefore: it's complicated. If one were to take proportional balance as a lens to view the world, the bias actually goes the other way, as smaller countries would appear favored. It is also only a global outlook, ignoring the functions of local or regional festivals (see also the Supplementary for region-by-region models).
Given the nature of this database, it is also difficult to ascertain whether the current distribution rises from festival organizers selecting relatively more films from smaller countries and less prosperous regions, or higher-resource countries not producing or submitting as many films as they could.
From yet another perspective, the film festival ecosystem as a whole could be seen as already managing to strike a middle ground between these two conceptualizations of equality --- whether by default or by design --- with different festivals screening different content and catering to different international, regional and thematic audiences \parencite[cf.][]{zemaityte_quantifying_2024}.

Figure \ref{fig_gdppop} provides three additional views into the dataset. Each country is shown as a marker, the size of which reflects the number of produced or co-produced film listings in festivals in the circuit. Reflecting the map in Fig. \ref{fig_map}, most of the film sources are indeed clustered on the ends of the scales in \ref{fig_gdppop}.A, with France and the USA the largest. 
The question remains, how well this particular festival database reflects country-wise film production capacity, and what could be said about the relative festival success of a given country without festival acceptance and rejection data. 
When compared with the UIS database (see Methods and materials), the correspondence might be close enough, depicted in \ref{fig_gdppop}.B. The range of the latter database is narrower, so the Cinando data are also cropped accordingly, but the positive correlation  (on log scales) is rather strong, especially given the disparity between the two databases discussed in Methods (adjusted $R^2=0.57$ i.e. 57\% variance in festival appearances is described by the production numbers). 
Most of the largest producers perform even better than expected relative to the others (points above the regression line). The largest countries in the world, China and India, appear both as relatively underrepresented, given their production volumes. As discussed in methods, it is difficult to tell at this point on these data alone (but would be interesting to know) whether this is because their filmmakers do not apply to that many festivals, if their films are not selected, or if they do appear in festivals but just not those recorded in the Cinando dataset. Not all films in all genres are of course produced with festivals in mind, and potential regional differences in this dimension may also affect these results.

\subsection{Predicting festival performance}

Due to the data-specific limitations and those of aggregation discussed in Methods, it is not possible to model festival selection processes directly here, but factors of representation based on country aggregates can be quantified. Figure \ref{fig_gdppop}.C depicts the residual values from a linear regression model predicting the 
($\log_{10}$) number of appearances of (co)productions in the circuit per country (henceforth simply referred to as N).
The predictors in the model \parencite[extending preceding research in][]{zemaityte_peripheral_nodate} are the population and GDP per capita of each country (centered at 10M and 10k, respectively), the number of festival events hosted in the country (with +1 Laplace-smoothing to allow for log-scaling), and distance from France in thousands of kilometers. All but the last are modeled on the $\log_{10}$ scale, with an interaction allowed between population and prosperity. The reasoning for the event count variable is that countries may be more successful if they host their own festivals, which may be more favorable towards local films. The distance variable reflects the observation that the festival circuit, as recorded in the database, is rather Euro-centric, which may provide advantages to nearby countries in terms of ease of travel
(often not required for participation but potentially supporting dissemination)
or cultural proximity \parencite[][]{straubhaar_beyond_1991,fu_examining_2010}. The model was checked for and met the assumptions of linearity, equal variance, and homoscedasticity, and exhibited only mild multicollinearity. Table \ref{table_reg} shows the coefficients of the model.
ü
\begin{table}[ht]
\centering
\begin{tabular}{lllll}
  \hline
 & $\beta$ & SE & $p$ \\ 
  \hline
Intercept & 2.05 & 0.08  & \textless0.001 \\ 
  Population & 0.72 & 0.06  & \textless0.001 \\ 
  GDP per capita & 0.84 & 0.08  & \textless0.001 \\ 
 Events & 0.29 & 0.12 & 0.02 \\ 
  Distance from France & -0.05 & 0.01 & \textless0.001 \\ 
  Population : GDP per capita & 0.19 & 0.07  & 0.01 \\ 
   \hline
\end{tabular}
\caption{Multiple linear regression model predicting film appearance counts by demographic, economic and geographic variables. Response and predictor variables, except for distance, on the $\log_{10}$ scale. $F(5,157)=105$, adjusted $R^2=0.763$.
} 
\label{table_reg}
\end{table}

All predictors are significant at the $\alpha=0.05$ level, and the model as a whole is highly predictive of the outcome, describing 76\% of the variance in country-wise festival representation.
All $\beta$ coefficients except distance are positive, indicating that, as already illustrated in the graphs above, larger and wealthier countries have on average more films appearing in the circuit, and also that hosting festivals indeed increases representation.
The intercept of $10^{2.05}=112.2$ corresponds to the expected N for (or a hypothetical mid-sized country at) the reference values of the variables, i.e. 10M population, \$10k per capita, hosting zero events, located where France is.
Since this is a log-log model, the coefficients are directly interpretable as percent changes (elasticities). A 1\% change in population (with everything else held constant) leads to a 0.72\% increase in predicted N; e.g. for a country of 10M, 1\% or 100k more people predicts 
$10^{\textrm{n}} - 10^{2.05}=0.8$ or almost one more appearance (where n is the predicted number of N appearances: $2.05+0.72 \times c$, where c is the 10M-centered target population $\log_{10}(1,1\textrm{M})-\log_{10}(1\textrm{M}))$. A million people more predicts about 8 more appearances.
However, their interaction is also significant, indicating that their effect is multiplicative: having both a large population and a high GDP per capita boosts festival appearances more than being just big or just wealthy. 

Distance from France is significant and negative: for each additional 1000 kilometers, N decreases by a multiplicative factor of $10^{-0.05}=0.89$, or in percent by $(1-0.89) \times 100=11\%$. Notably, this factor is significant even in this model that controls for GDP per capita (so it is not just an effect of many less prosperous countries e.g. the Global South being far from Europe). A simpler model just predicting N by distance describes about 7\% in variance. The fact that even this crude proxy to geographical effects holds predictive power indicates that where a country is located may significantly affect its chances of festival representation.

In summary, most of the variation in how many films 
get selected 
from each country comes down to their size, wealth, where they are relative to Europe, and how many festivals they host themselves.
While this is somewhat obvious, a statistical model like this allows for a more precise quantification of that intuition.

Coming back to Figure \ref{fig_gdppop}.C: the residuals of this model indicate which countries are near their predicted values (those around zero), which do better than average, and which do worse. The zero is effectively the regression line and can be thought of as the global average prediction. The residuals are on the same $\log_{10}$ scale: a value of 1 means the country appears 10x more in the circuit than would be expected by the aforementioned factors. 
For example, Argentina, despite its moderate average GDP per capita of \$12k and about 44M population in the period, has 2041 
film-festival pairs
listed in the database, while the global model prediction would place it at 290. It is shown below how this may be boosted by a local festival circuit \parencite[see also][]{isaza_branding_2012}. 
The over 10 times smaller Uruguay still has 270 appearances of 142 films on record (prediction would be just 23).
France in comparison performs slightly lower than predicted (has 6759, prediction is 6541), despite hosting a large array of festivals. The USA and China have considerably lower N than would be expected by their size and prosperity. As discussed above, this may be an effect of local markets and festivals. 
The UK is shown to under-perform despite its prosperity and arguably favorable position in European audiovisual industries, especially in the pre-Brexit era that is part of the observation period here \parencite{donders_decline_2016,steemers_international_2016,ibrus_quantifying_2023}.
Afghanistan is among countries with a rather high positive residual: despite its very low GDP and other factors, it is listed on 47 appearances of 26 films in total in 2012-2021, although 23 of them appear to be co-productions with mostly European countries. Many small European countries also beat the odds, including Iceland and Luxembourg, but also Estonia, Latvia and Lithuania (all below 3 million in size), potentially reflecting recent developments of their film and television industries \parencite[cf.][]{hjort_cinema_2007,iosifidis_public_2007,ibrus_searching_2019,ibrus_quantifying_2023}.

\subsection{Cultural flows of festival films between countries of the world}

These results raise the question of how films may travel between countries and festivals, how much apparent success comes down to (potentially favorable) local festivals, and how much countries may differ in terms of cultural "trade balance" \parencite{disdier_bilateral_2010,baorui_currencies_2024} --- in this case, how much a country's films get exported to foreign festivals, versus locally hosted festivals "importing" foreign films.
These questions have been explored in the past in terms of television or cinema, but not festivals, and usually on a country rather than world scale \parencite{meloni_american_2018,mckenzie_blockbusters_2018,crane_cultural_2018,coate_feature_2017,avezzu_market_2022,ibrus_quantifying_2023,weber_global_2024,holobut_statistical_2024}.
While all these topics deserve deeper study, this section attempts to shed some light on these global dynamics and cultural flows \parencite[see also][]{disdier_bilateral_2010,fu_examining_2010,christensen_transnational_2013,kawashima_asian_2018,hartley_digital_2020}.
This analysis is of course still limited to festivals and does not consider other venues and platforms. 

This requires once again choosing between potential aggregation approaches. In the analyses above, every 
film-festival entry
of a film produced or co-produced by a country was counted equally for simplicity. This is fitting if one is interested in whether a country is represented in a festival or the circuit as such. In this section, a slightly more nuanced approach is taken, and the counts and shares are weighted by the number of co-producers, for a hopefully more accurate picture of cultural flows. The unit is still one film 
entry
at a festival. For example, a film produced by country A and shown in a festival in country X adds a count of 1 in the direction from A to X; while a film co-produced by countries A and B adds 0.5 for each instead.

\begin{figure}[htb]
	\noindent
	\includegraphics[width=\columnwidth]{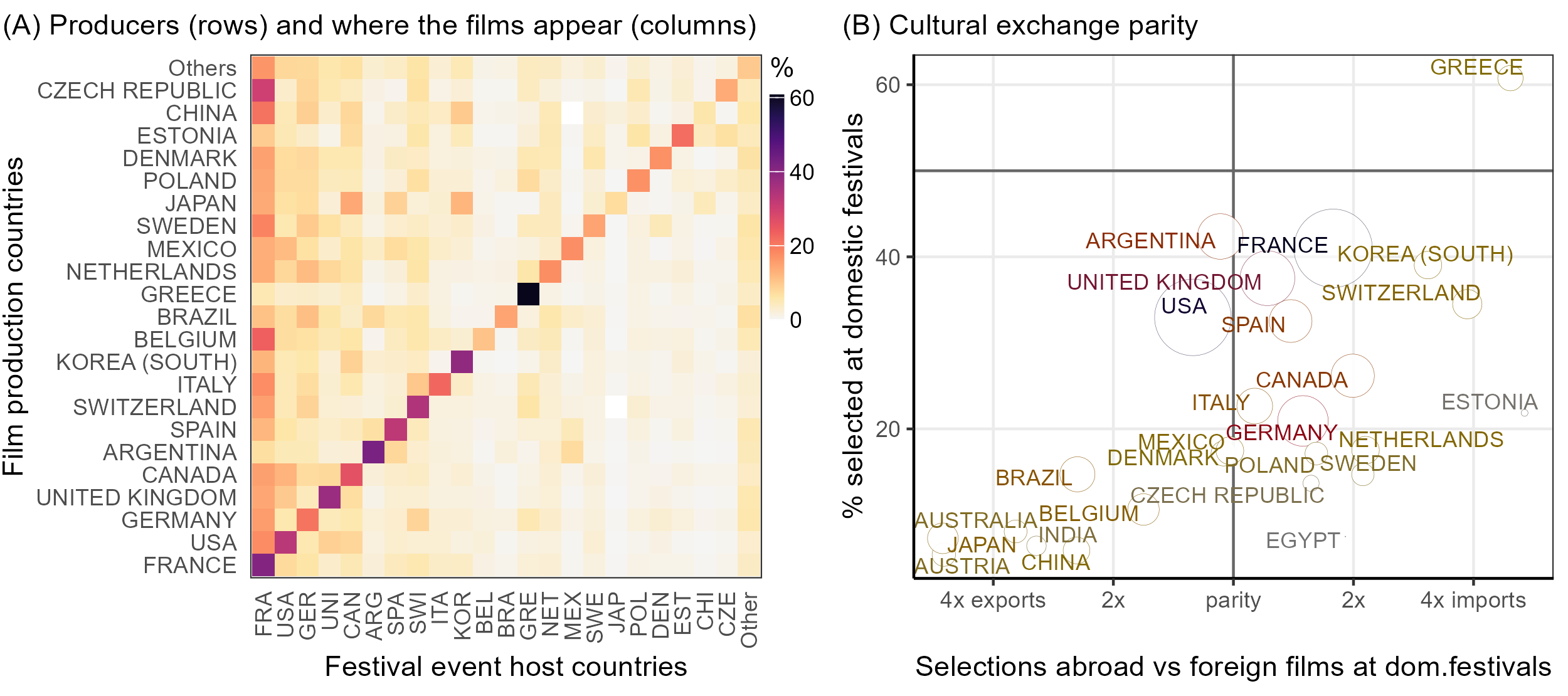}
	\caption{
 The flow of films between production countries and festival-hosting countries. Panel A depicts the export share of festival films normalized by production countries (rows). Panel B displays the export-import balance on the horizontal axis on a $\log_2$ or multiplicative scale (not taking into account domestic screenings of domestic films), against the share of domestic film selections at domestic festivals on the vertical axis, for all countries with at least 5 hosted festival events in the database. The largest producer, France, also imports more than it exports, i.e. more foreign films occur in its festivals than its own films are 
selected
 at festivals abroad.
	}\label{fig_flows}
\end{figure}

Figure \ref{fig_flows}.A illustrates the flow of films between the largest production and festival host countries. Not all those listed are large producers, for example, Estonia once again appears because of its Tallinn Black Nights film festival, which screens a fairly large number of films yearly. Normalized by rows (producers), the diagonal of the matrix effectively displays the share of a given country's productions that are screened locally. For some countries this is quite high: 61\% for Greece (at two festivals, the Thessaloniki Documentary Festival and the Thessaloniki IFF). For Argentina and France, this is 42 and 41\%. For others like China and Japan (and on average other countries) domestic 
festival selections of films
are at or below 10\%. Some countries export a considerable share of their films to a single destination. A fourth of Belgian film entries are at French festivals --- not surprising given the geographic proximity and one shared national language). Some relationships are reciprocal: the UK screens almost a tenth of their films in the USA and vice versa, although the primary non-domestic destination for US films is France (18\%). The latter is also an important destination for the UK at 14\% (or 311 entries), but interestingly not so much the other way around, with only 3\% or 143 French film listings in UK festivals \parencite[see also][]{mazdon_je_2010}. A larger version of this graph may be found in the Supplementary.

Not all countries in the database are recorded as hosting any festivals, which means their incoming number of films here is zero. Figure \ref{fig_flows}.B shows a sample of countries that have hosted (at least 5) festival events, arranged across an export-import parity axis. This is calculated simply as a multiplicative difference or $\log_{2}(\textrm{imports}/\textrm{exports})$ (where e.g. a value of 1 means 2x more imports than exports, 2 is 4x more, etc). Here imports are defined for each country as the sum of (weighted) 
film-festival pairs 
where the producer is not the given country, and exports as the sum of instances where a given country is a producer but not the host. The vertical axis shows how many entries of a country's films occur at their own festivals (effectively the diagonal of the left side panel). Smaller production countries like Estonia and Sweden import more than they export. Despite its large production volume, 1.8x more foreign films are listed at French festivals, compared to how many French films are selected at festivals abroad. A reason may be its relatively large internal festival circuit (76 events in 19 series in 2012-2021). Along with other large production countries like the UK, USA, and Argentina, a fair share of French films are screened within the country. 
The USA is a net exporter not only of cinema but also festival films \parencite[cf.][]{meloni_american_2018,tunstall_media_2008}.
The largest countries China and India are in the bottom left, exporting 2.5 and 3.1x more, and screening their productions largely in festivals abroad. This is however according to the somewhat Euro-centric Cinando database, which for example lists only the Shanghai and Beijing IFF as Chinese festivals. It is entirely possible that while some films from these countries occur in Western festivals and get listed in the database, the share of local festivals may simply not be represented here \parencite[cf.][]{ma_regarding_2014,dastidar_indian_2020}. 

\begin{figure}[htb]
	\noindent
	\includegraphics[width=\columnwidth]{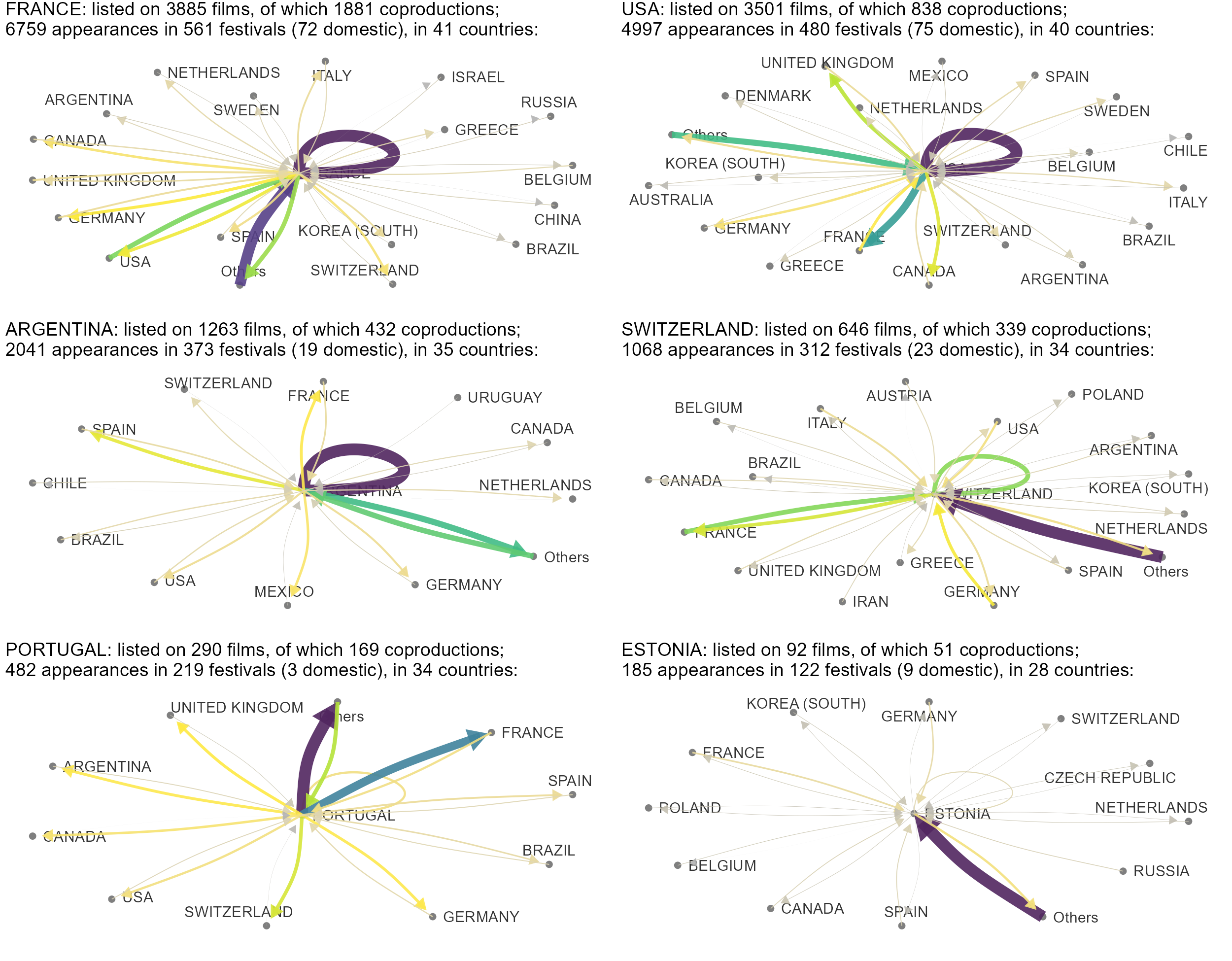}
	\caption{Film exchange flows of selected countries. These "star networks" only show links between a given central country and countries where its films are showcased at festivals (outgoing arrows) or whose films are shown in its festivals. The line color and size are relative, with the thickest being the largest link in terms of either in- or outgoing film appearances. To maintain legibility, only those countries accounting for about a fifth of the imports-exports are shown for each country, the rest grouped as "Others". As already seen above, some countries like France have strong internal festival circuits, visible as the loops in the network, while others export either in a focused or spread-out manner.
	}\label{fig_nets}
\end{figure}

Figure \ref{fig_nets} further illustrates the flow of festival films for a selection of countries, ranging again from large producers like France and the USA to small players like Estonia and Portugal, which also rely more on international co-productions. The data is aggregated as above. These networks only show the inflow and outflow for each target country at a time. With the arrow size indicating relative counts, one this that is immediately striking is the difference in the sizes of the local circuits, visible as the looping edge. As also shown in Fig. \ref{fig_flows}.A, countries also differ in their cultural trade partner distributions, with some sending and receiving films from a variety of nations while others are somewhat more focused on certain destinations like the UK to France. Some links here are also due to co-productions, e.g. a film jointly produced by Portugal and Spain, and screened at the San Sebastian Film Festival in the latter, would still count (at a 0.5 weight) towards a Portugal-to-Spain link. 

It should also be noted that this snapshot of the Cinando database ends in 2021, therefore not reflecting the state of the festival industry since then, including the banning of Russia from major film festivals following its military invasion into Ukraine in 2022 \parencite{vivarelli_berlin_2023}, or the defunding of the Argentinian national film body in 2024 \parencite{geisinger_argentinian_2024}.

\subsection{How much diversity is gained by including films from smaller countries?}

As illustrated in the beginning in Figure \ref{fig_map}, a handful of the largest production countries make up the lion's share of the 
film entries
across festivals. Yet the circuit is not as dominated by large and prosperous countries as it could be if representation was entirely proportional to population or the offerings in perfect correlation with relative GDP (Fig. \ref{fig_violins}). It has been shown in recent research that festivals vary significantly in terms of diversity along multiple dimensions, including gender balance, and thematic, linguistic, and geographic diversity \parencite{zemaityte_quantifying_2024}. While some are regional or focus on films from one or a few countries, others are highly international. There are thematically concentrated festivals like Frightfest London, Fantasia, and DocCorner; festivals that screen films predominantly in one language like Spanish in Latin America (Guadalajara, Ventana Sur) or Sundance in English, yet others where no single language takes up much more than 10\% (Tallinn Black Nights in 2019). Similar variation can be found in regional versus international representation. And when it comes to gender, some festivals like Göteborg or Hot Docs often have an almost equal balance, while others can be male-dominated in some years to the point of featuring no women directors at all \parencite[see][]{zemaityte_quantifying_2024,verhoeven_re-distributing_2019,ehrich_film_2022}.

A question that has been raised is how much diversity (and potential public value) is added to cultural and audiovisual spheres, including the film festival circuit, by including products of smaller or peripheral countries
\parencite{sand_small_2019,zemaityte_quantifying_2024,ibrus_quantifying_2023,felix-jager_effects_2020,zemaityte_peripheral_nodate}.
While it would be difficult to quantify the exact contribution of individual countries, it is possible to simulate what the circuit would look like \emph{without} certain countries, and roughly interpolate their estimated contribution from that. For example, if removing a group of countries from the dataset does not induce a change in diversity it can be surmised that their contribution to the given dimension of diversity was negligible. See Methods for details about the simulation, diversity calculation, and bootstrapping. As in the previous section, if a film has multiple tags (here languages), their contribution is weighted so they sum to 1. Genres are allowed to be compositional averaged so that each film is represented by one thematic vector \parencite[as in][]{zemaityte_quantifying_2024}.

\begin{figure}[htb]
	\noindent
	\includegraphics[width=\columnwidth]{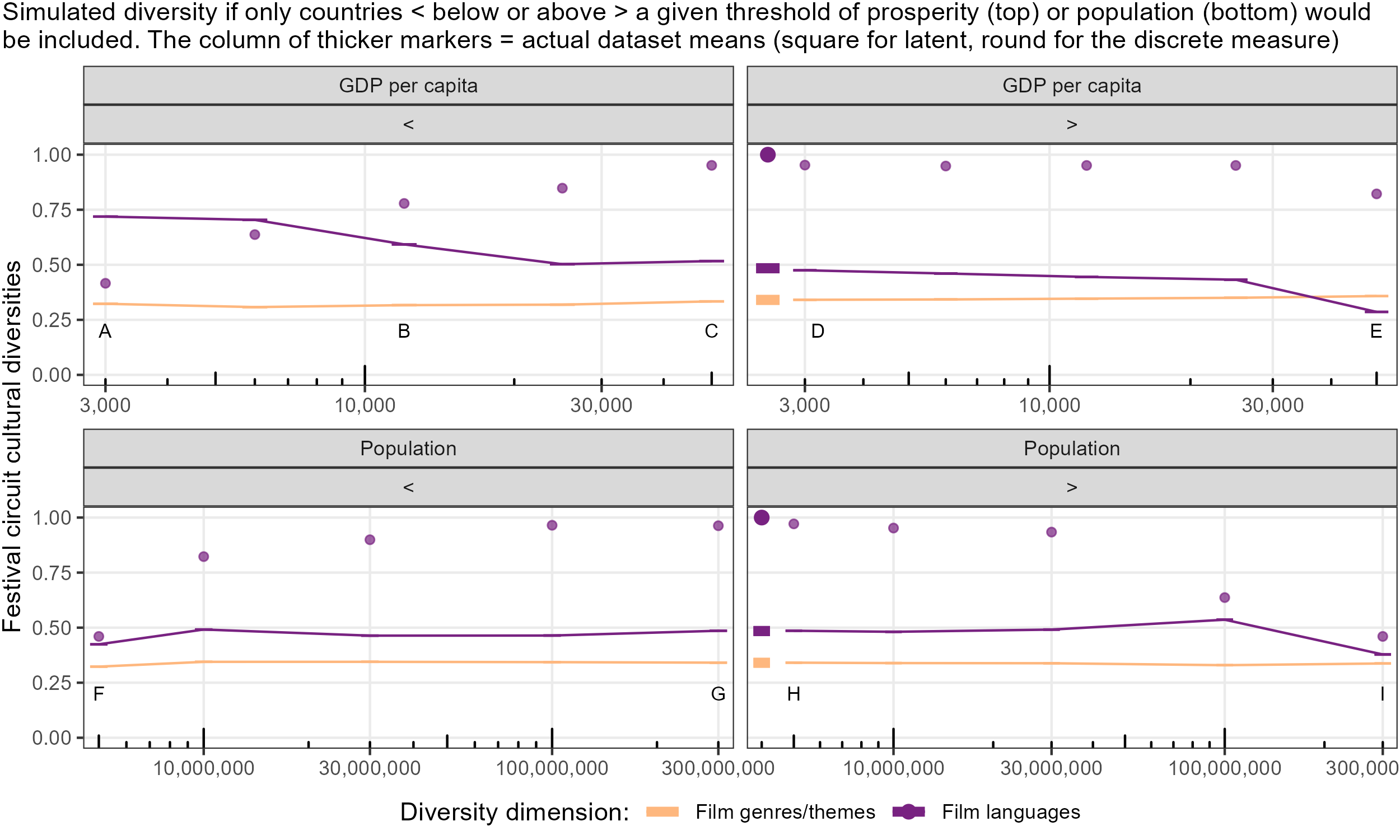}
	\caption{Assessing the contribution to diversity by countries of variable socioeconomic profiles via simulation. On the far left side of both the GDP and population panel, the hypothetical circuit consists of only countries with the smallest respective values. The middle shows circuits of countries below $<$ or above $>$ a given threshold, and the right stands for circuits of only the largest and wealthiest. The purple dots stand for linguistic diversity as counts of languages (percent of the actual value), and the lines are the latent continuous measures of linguistic (purple) and thematic diversity (orange). The thicker bars and dots in the middle mark the averages of the actual festival circuit. The graph also displays 95\% confidence intervals (via normal approximation), displayed as error bars around the latent measure points, but these are so narrow as to be barely visible, indicating the results are quite stable.
   The graph is annotated with letters illustrating the samples as follows, starting with the top left.
    A: only the 47 poorest countries that have participated in the circuit, with mean GDP per capita below 3k (e.g. Syria, Cambodia, Laos). B: the former as well as countries like Kazakhstan, Mexico, China, Belarus. C: all countries except those above 50k (so excluding Sweden, USA, Singapore, Switzerland, etc.). D: all except the poorest (below 3k). E: only the wealthiest (above 50k). Bottom row, F: only the 50 smallest countries (below 5M population). G: all except the largest (USA, India, China). H: all except the aforementioned smallest. I: only the largest three.
	}\label{fig_simu}
\end{figure}

Figure \ref{fig_simu} illustrates the simulation results: for both GDP per capita (top) and population (bottom), each marked position depicts the re-calculated diversity after removing larger and wealthier countries (left) and progressively the smaller and less wealthy countries (right). Neither dimension is of course a direct cultural diversity measure as such, but can be considered as proxies or rough estimates of cultural representation. The figure shows the continuous diversity measure (lines) and discrete counts for languages (as a percent of the actual circuit total; the averaged genres cannot be directly counted). Points of interest discussed in the text and caption have been marked with letters A to I.

To break these results down: first of all, thematic diversity is practically unaffected and remains the same flat line regardless of the composition of the hypothetical festival circuit scenarios. Removing small or less prosperous countries from the circuit (pointers D, H) would not affect thematic diversity, but neither would do the opposite (pointers A, F). This indicates that on average, a large set of small countries can produce as much thematic variation as a small group of the largest ones. This may be somewhat of an artifact of the data, however: about a fourth of all entries contain "drama" as a thematic tag anyway (closely followed by "documentary"), with the other tags much less frequent each. A more thorough analysis of actual film content would likely provide more interesting results.

Linguistic diversity is slightly more complex. A circuit comprising only the 47 poorest countries (pointer A, top left in Fig. \ref{fig_simu}) has only about half the current languages represented, but the continuous (latent vectors-based) diversity measure is slightly higher, reflecting the high typological dissimilarity of languages outside of wealthy Western countries. However, removing all these countries from the circuit (D) barely affects its global diversity average. The circuit only becomes somewhat less diverse past the \$50k mark (E), i.e. if it would consist of only the wealthiest nations like Singapore, USA, etc. This is somewhat surprising, but explainable by the high number of languages present in the diverse productions of these large film industries. It is also not impossible that some film entries may lack the full list of co-production countries, while the languages spoken in them are listed.

When it comes to population, removing larger countries (bottom left, A) has again little effect on latent linguistic diversity, and the counts fall similarly as for GDP. 
The other way around, removing the 50 smallest countries of up to 5M population --- including countries like Iceland, Estonia, Luxembourg, and Uruguay --- makes little difference (see H). 
One reason is that this is not a comparison against a model of fair or uniform representation, but the current state of affairs: their share in the current circuit is already quite small, so removing them does not change that much.
A hypothetical circuit consisting of only (contributions proportional to the current circuit from) the USA, India, and China (bottom right, I), would be dominated by English, and have less than half (52) of the current languages present. Yet, the latent linguistic diversity would fall only marginally, as the share of otherwise frequent and similar Indo-European languages like French and Spanish would fall. Still, a fairly large number of diverse languages would be present.

In summary, when it comes to cultural diversity --- as represented by languages spoken in films and their thematic genre --- small countries \emph{currently} add very little from a purely global position.
They however make up only a small part of the circuit as of now, while a hypothetical circuit of just smaller countries would be as or almost as diverse along these two dimensions, compared to current reality. The interpretation is, therefore, that smaller countries may harbor untapped potential to increase linguistic and cultural diversity in the festival ecosystem.
Naturally, such a simplified global aggregation neither is nor should be the only perspective to diversity, and festivals often serve many (if not more) local cultural, artistic economic, and other functions discussed in the Introduction.

\section{Discussion}

This contribution has attempted to provide a global-scale quantification of national representation in the film festival circuit from an (admittedly simplified) demographic and economic perspective and to explore the dynamics and cultural flows shaping the festival network, which in turn impacts the economies and cultural landscapes of the production and event host countries.
Rather than conclusive outcomes, these results and methodological proposals are meant more as a stepping stone towards deeper and more contextualized research of the festival ecosystem and cultural value generation, building on and complementing recent investigations of the festival circuit via the Cinando database \parencite{zemaityte_quantifying_2024,zemaityte_peripheral_nodate}.
Several simplifications were necessary due to the nature of the data, and the subsequent limitations should be kept in mind when interpreting the results. These aspects can also hopefully be improved upon in the future.

For example, the diversity simulation in the last Results subsection is of course a very coarse first analysis, as cultural diversity (and public value) has many more dimensions. The global averaging also simplifies and overlooks the value that regional festivals can provide by platforming local or regional films, as well as the (both symbolic and economic) value that emerging industries and filmmakers can gain from being showcased at prestigious international film festivals \parencite{hadida_commercial_2009}. 
It is hoped however that this methodological proposal paves the way for a more systematic and data-driven approach to cultural diversity and relatedly, public value, a domain which has until recently remained rather theoretical in the context of film festival and audiovisual industries research \parencite[but see][]{faulkner_avoiding_2018,ibrus_quantifying_2023,zemaityte_quantifying_2024,zemaityte_peripheral_nodate}.

An aspect touched upon here but deserving of further study concerns the patterns and dynamics of co-productions \parencite{bondebjerg_transnational_2016,parc_understanding_2020}. Many of the country examples brought here exhibit high rates of international production, with some co-producing half or more of their festival films (see Figure \ref{fig_nets}). The production patterns are also likely predictive of cultural flows, as films produced jointly may well subsequently enjoy reception at festivals in all involved co-producer countries. 

Beyond the bias analysis, there was also little attention given here to the categorization of festivals as the accredited A-list and the rest, which of course comes into play in representation and success, given the known variance in competitiveness and functions of these events \parencite{zemaityte_quantifying_2024,zemaityte_peripheral_nodate}. A related major limitation imposed by the nature of the dataset is the lack of information on rejection. While the Cinando database provides details on festival programming, there was no way to tell how many films a given country actually sends to a given festival, nor the percentage that gets selected, and whether there is a bias towards or against some countries. Obtaining this information would greatly support the understanding of both the dynamics of individual festivals and the circuit as a whole. It would also enable a more informative modeling of representation and its various potential predictors, which was done here in a coarse manner, simply counting 
film-festival pairs 
per country. Future research could also incorporate more precise information on film industry revenues instead of country GDP, or factors of cultural proximity between producer and host countries \parencite{fu_examining_2010}. 
The latent embedding approach proposed in \textcite{zemaityte_quantifying_2024} is suitable for computing both diversity and similarity, in a manner that goes beyond discrete counts of cultural markers or categories and allows for taking into account their intra-similarity.
The diversity analysis above considers thematic genre, but further cultural and content factors as well as the perceived quality and novelty of a film likely affect its success at festivals, and likely interpersonal, professional network, and personal prestige-based factors and biases too \parencite{de_valck_fostering_2016,coate_feature_2017,mair_role_2020,ehrich_film_2022}.

It is nevertheless hoped that these analyses and methodological considerations provide support for further quantitative research into film festivals. The preliminary insights could be useful for researchers of the topic but perhaps also for festival organizers and policymakers, providing a data-driven perspective and complementary ways of thinking about analyzing film festival programming, their socioeconomic aspects, as well as diversity and balance.

\section{Conclusions}

Film festivals boost economies, provide a platform for filmmakers to showcase their work, enhance cultural spheres and promote international interactions, provide a spotlight on underrepresented voices and narratives through selective programming, and directly or indirectly create public value. At the same time, the global festival circuit, as represented in the Cinando database, is dominated by a few successful production countries. This contribution has endeavored to measure this apparent bias, discuss quantifiable models of representation equality, show that representation at festivals is highly predictable by socioeconomic factors, and explore the global cultural flows of festival films. While the circuit is indeed saturated by films from large affluent Western nations, it also platforms smaller countries, which have the potential to increase cultural diversity in the festival circuit.

\section*{Acknowledgements}

A.K. was supported by the CUDAN ERA Chair project for Cultural Data Analytics, funded through the European Union Horizon 2020 research and innovation program (Project No. 810961). 
V.Z. was supported by the Public Value of Open Cultural Data project, funded through the Estonian Research Council (Project No. PRG 1641), and by project CresCine - Increasing the International Competitiveness of the Film Industry in Small European Markets, funded through the European Union Horizon Europe funding scheme (Project No. 101094988). The funders had no role in study design, data collection and analysis, decision to publish, or preparation of the manuscript.

\section*{Author contributions}

A.K. designed the research, prepared the data, designed and performed the data analysis, wrote the text, and created the figures. V.Z. suggested the initial research direction of sections 3.1-3.2, including the WB and FIAPF data, and provided comments on the final draft.

\section*{Data availability}

The Cinando database snapshot is available via \url{https://doi.org/10.6084/m9.figshare.22682794.v1} (Cinando technical ID values are anonymized). 
The data are also available to explore via an interactive dashboard that accompanied the first Cinando-related publication: \\ \url{https://fiapf.org/festivals/accredited-festivals/competitive-feature-film-festivals}.
The {FIAPF} accreditation list is available via \url{https://fiapf.org/festivals/accredited-festivals}.
The population and GDP data were retrieved from the World Bank database available via 
 \url{https://data.worldbank.org/indicator}.
The UIS film production statistics are available via \url{https://data.uis.unesco.org}.

\printbibliography

\FloatBarrier
\newpage

\section*{Supplementary Materials: additional graphs}

This appendix contains a few more graphs, expanding the content that was explored in the main text.

\begin{figure}[htb]
	\noindent
	\includegraphics[width=\columnwidth]{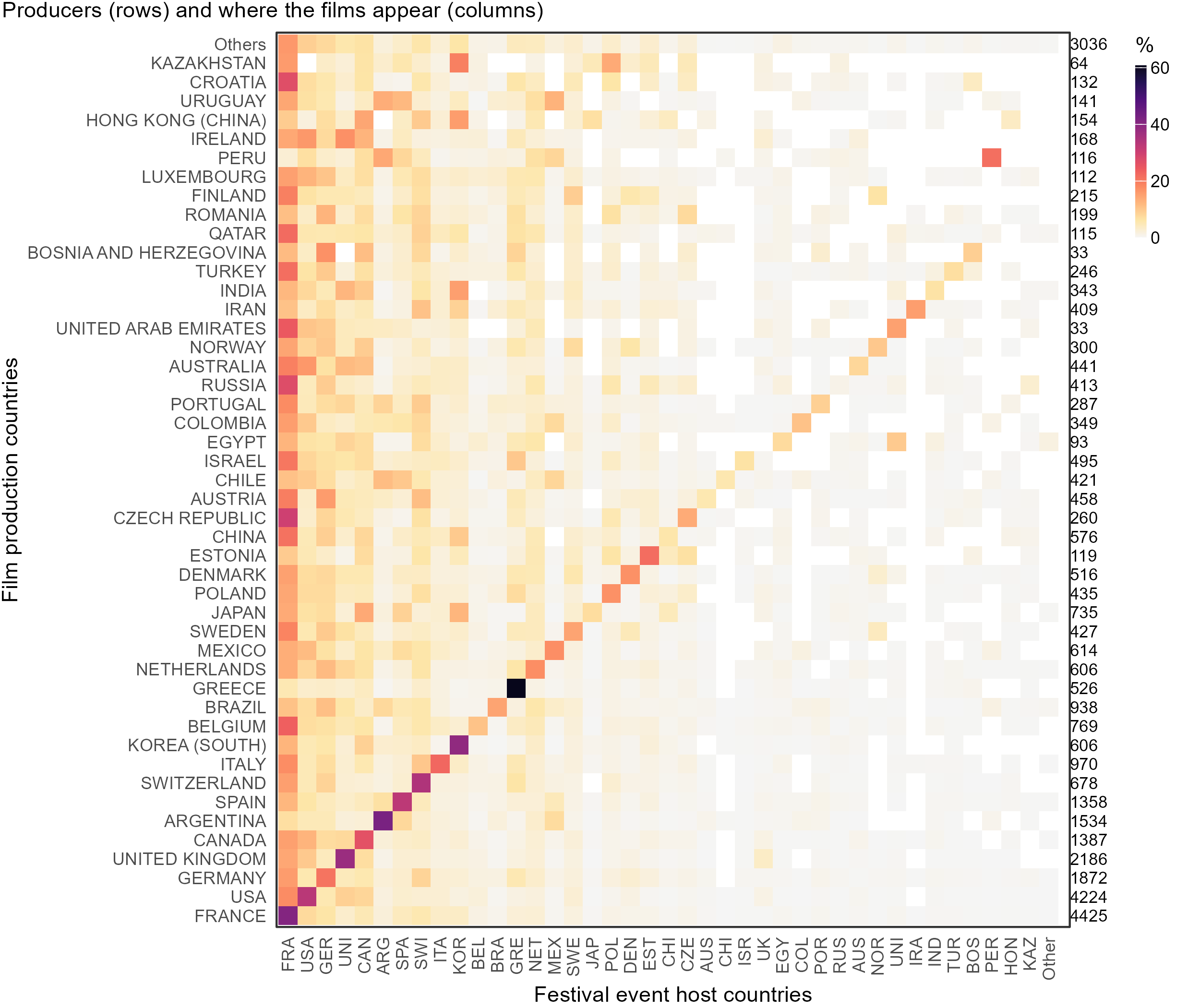}
	\caption{This graph complements the flows matrix in the main text, showing more countries, and the total (rounded) sums of production by the production countries (right side numbers column). The values differ slightly from the absolute values of database entry sums discussed in the text, being based on the co-production weighted values, as in the flow analysis. As in the main figure in the text, rows are producers and columns are host countries, and the cells, normalized row-wise, show the percentage of %
film-festival pairs 
in each host country.
	}\label{fig_sup_flows}
\end{figure}

\vspace*{3in}

\begin{figure}[t]
	\noindent
	\includegraphics[width=\columnwidth]{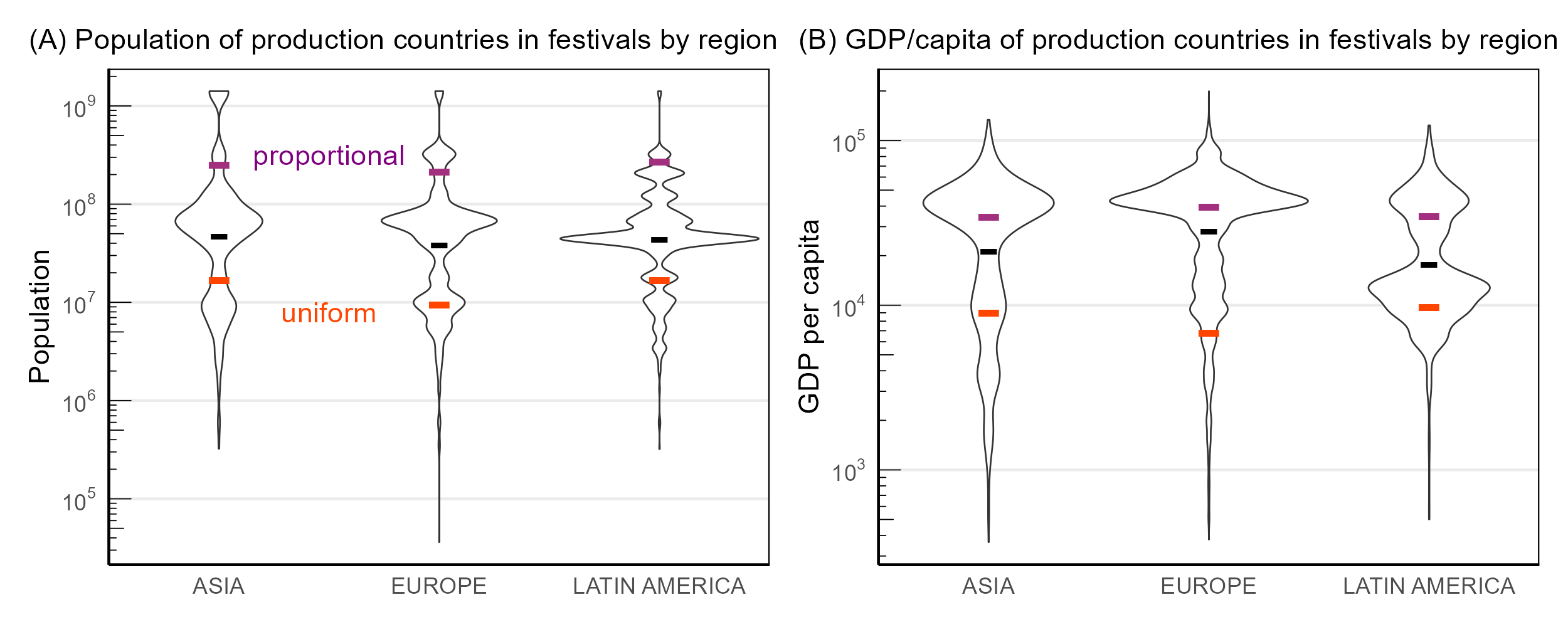}
	\caption{This graph complements the global distributions graph in the main text, showing the distributions and means across three regions, showing the average population (A) and GDP per capita (B) of films in festivals hosted in these regions (films coming from other regions are included).
	}\label{fig_sup_extraviolins}
\end{figure}

\end{document}